\documentclass[journal=ancac3,manuscript=article]{achemso}

\usepackage[version=3]{mhchem} 

\usepackage[usenames,dvipsnames]{xcolor}

\DeclareUnicodeCharacter{2212}{-}

\newcommand{\est}{[Au-TiO$_{2}$-CO] }

\newcommand{\tio}{TiO$_{2}$ }

\author{Matias Berdakin}
\email{matiasberdakin@unc.edu.ar}
\affiliation{INFIQC (CONICET-UNC), Ciudad Universitaria, Pabellón Argentina, 5000 Córdoba, Argentina.}

\alsoaffiliation{Departamento de Química Teórica y Computacional, Fac. de Ciencias Químicas, Universidad Nacional de Córdoba, Ciudad Universitaria, Pabellón Argentina, X5000HUA Córdoba, Argentina.}
\author{German Soldano}
\affiliation{INFIQC (CONICET-UNC), Ciudad Universitaria, Pabellón Argentina, 5000 Córdoba, Argentina.}

\alsoaffiliation{Departamento de Química Teórica y Computacional, Fac. de Ciencias Químicas, Universidad Nacional de Córdoba, Ciudad Universitaria, Pabellón Argentina, X5000HUA Córdoba, Argentina.}
\author{Franco P. Bonafé}
\affiliation{Max Planck Institute for the Structure and Dynamics of Matter, Center for Free Electron Laser Science, Hamburg, Germany.}
\author{Varlamova Liubov}
\affiliation{Bremen Center for Computational Materials Science, Universitát Bremen, Bremen, Germany}
\author{Bálint Aradi}
\affiliation{Bremen Center for Computational Materials Science, Universitát Bremen, Bremen, Germany}
\author{Thomas Frauenheim}
\affiliation{Bremen Center for Computational Materials Science, Universitát Bremen, Bremen, Germany}
\alsoaffiliation{Computational Science Research Center (CSRC) Beijing and Computational Science and Applied Research (CSAR) Institute, Shenzhen, China}
\author{Cristián G. Sánchez}
\affiliation{Instituto Interdisciplinario de Ciencias Básicas, Facultad de Ciencias Exactas y Naturales, Universidad Nacional de Cuyo, CONICET, Padre Jorge Contreras 1300, Mendoza M5502JMA, Argentina}

\title[An \textsf{achemso} demo]
  {Dynamical evolution of the Schottky barrier as a determinant contribution to electron-hole pair stabilization and photocatalysis of plasmon-induced hot carriers}

\keywords{ Plasmon-induced hot carriers, Hot carrier stabilization, 
Hot carrier mediated catalysis, Real-time electron-ion quantum dynamics, Plasmonics }

\begin{document}

\begin{tocentry}

Some journals require a graphical entry for the Table of Contents.
This should be laid out ``print ready'' so that the sizing of the
text is correct.

Inside the \texttt{tocentry} environment, the font used is Helvetica
8\,pt, as required by \emph{Journal of the American Chemical
Society}.

The surrounding frame is 9\,cm by 3.5\,cm, which is the maximum
permitted for  \emph{Journal of the American Chemical Society}
graphical table of content entries. The box will not resize if the
content is too big: instead it will overflow the edge of the box.

This box and the associated title will always be printed on a
separate page at the end of the document.

\end{tocentry}

\begin{abstract}
The harnessing of plasmon-induced hot carriers promises to open new avenues for the development of clean energies and chemical catalysis. The extraction of carriers before thermalization and recombination is of primordial importance to obtain appealing conversion yields. Here, hot carrier injection in the paradigmatic Au-TiO$_{2}$ system is studied by means of electronic and electron-ion dynamics. Our results show that pure electronic features (without considering many-body interactions or dissipation to the environment) contribute to the electron-hole separation stability. These results reveal the existence of a dynamic contribution to the interfacial potential barrier (Schottky barrier) that arises at the charge injection pace, impeding electronic back transfer. Furthermore, we show that this charge separation stabilization provides the time needed for the charge to leak to capping molecules placed over the TiO$_{2}$ surface triggering a coherent bond oscillation that will lead to a photocatalytic dissociation. We expect that our results will add new perspectives to the interpretation of the already detected long-lived hot carrier lifetimes, their catalytical effect, and concomitantly to their technological applications.
\end{abstract}

\section{Introduction}
Plasmonic nanoparticles (NP) have striking light-harvesting capabilities that stem from the collective oscillation of conduction band electrons in resonance with light electric field, process known as localized surface plasmon resonance (LSPR).\cite{2011Hartland} Harnessing this optical feature has triggered a wide stream of research in areas related to light conversion as photocatalysis, and the development of photovoltaic systems.\cite{Hartland2017,Linic2018} For small NPs Landau damping, the decoherence of electrons involved in the plasmon excitation, is the fastest dissipative mechanism of the energy absorbed by the LSPR photoexcitation, leading to the formation of electron-hole pairs excited beyond the thermal energy, usually named as hot carriers.\cite{Clavero2014,Nordlander2015,Govorov2014} 

For many years, it was a common understanding that the electron-electron scattering would hindrance the harnessing of hot carriers generated during the LSPR excitation. Nevertheless, in the last decade experimental evidence showed that hot carriers injection to the medium, capping molecules\cite{Boerigter2016} or semiconductors,\cite{Furube-jacs} take place by an ultrafast mechanism (lifetime of $\sim$50 fs) that competes with thermalization. This fascinating discovery has pervaded the nanoscience community and placed the plasmon-induced hot carrier generation and exploitation as a research mainstream.

Chemical catalysis driven by plasmon-induced hot carriers has been proven in several chemical reactions involving oxidation or reduction of small molecules,\cite{Christopher2011,Zhao_2018} degradation of organic molecules,\cite{Kochuveedu2012} and even in high energy cost reactions like water splitting\cite{mubeen2013,robatjazi2015direct,lee2012plasmonic} and H$_2$ dissociation.\cite{Halas2013} On the other hand, ultrafast injection to semiconductors has also been thoroughly studied experimentally.\cite{Hattori2019,wu2015efficient,du2013ultrafast} Furthermore, the lifetime of the injected carriers in semiconductors is under vigorous research because it sets the time bounds for their exploitation. Remarkably, this has been estimated in the order of hundreds of picoseconds\cite{catal10080916,time-resolved-xray} to nanoseconds,\cite{capturing_long_lived} challenging our interpretation of the problem and raising questions regarding the chemical catalysis mechanisms.\cite{Wei2018} On the side of theoretical research, much effort has been invested to disentangle the hot-carrier generation process and their energy distribution in several plasmonic materials by means of many-body,\cite{super-hot1,super-hot2,super-hot3,castellanos2019single} DFT,\cite{Govorov2014,Govorov2015} TD-DFT\cite{Erhart_acsnano}, DFTB\cite{Oscar2016,Oscar2019,interplay2019} and phenomenological master equation techniques\cite{Nordlander2014,Lischner_dband,Lischner_materials}. Interested readers can find comprehensive reviews of the latest experimental and theoretical advances in references \citenum{BESTEIRO2019,Chang_Le2019,Wei2018,Linic2018,Hartland2017,Brongersma2015}. 

Nevertheless, much less has been done in the field of ab-initio simulations of the charge injection dynamics and the subsequent catalytic effect of hot carriers. Some recent benchmarks along this line are worth highlighting. Hot carrier injection from a metallic cluster to semiconductors have been studied by nonadiabatic molecular dynamics,\cite{nano_Lischner,Prezhdo_instantaneus} and recently a detailed description of the charge localization in a Ag-TiO$_2$ hybrid system has been performed by means of real-time time-dependent DFT (RT-TD-DFT).\cite{ACSgao} Besides, nonadiabatic molecular dynamics has been employed to study the  dissociation of H$_{2}$ on gold NPs dimers,\cite{Nelson_acsnano} N$_{2}$ dissociation by hot carriers produced in silver wires\cite{Aikens_N2} and water splitting over gold NP\cite{Sheng_ACS}. In this context, by means of real-time time-dependent tight-binding DFT (RT-TD-DFTB) electronic and electron-ion dynamics, we studied the hot carrier injection in the paradigmatic Au-TiO$_2$ system. To simulate nanosized structures, a new DFTB parameterization is presented that successfully describes the electronic properties of the hybrid NP. Regarding the charge injection, our simulations show that pure electronic features (without considering many-body interactions or dissipation to the environment) are enough to justify the electron-hole separation stability. Typically, the prolongation of the hot-carrier lifetimes in metal-semiconductor hybrid systems is attributed to the Schottky barrier (SB), i.e. the potential asymmetry rendered by the charge accumulation at the interface that occurs due to the Fermi level equalization, and the cooling of the electronic subsystem that together with the SB locks the electronic backscattering. Our results show intrinsic electron-hole separation stability without cooling interactions and even in the presence of laser pumping. This unveiled the existence of a dynamic contribution to the Schottky potential barrier (DSB) that arises at the charge injection pace, locking the electronic back transfer. It's important to state that this mechanism differs from the dynamical detuning of acceptor-donor states described in reference \citenum{charly_trap}. This dynamic contribution to the electron-hole stabilization has been overlooked and can be a missing piece in the complex puzzle that renders the hot carrier lifetime in nanosized structures. Furthermore, the electron-hole stabilization provided by the DSB is important to describe the photocatalytic effect of the plasmon-induced hot carriers. Here we show that charge leaks to capping molecules placed over the TiO$_{2}$ surface triggering a coherent oscillation of the nuclei that will lead to a photocatalytic dissociation. We expect that our results will add new perspectives to the interpretation of the already detected long-lived hot carrier lifetimes, their catalytical effect, and concomitantly to their technological applications.

\section{Results and discussion}

\begin{figure}
 \centering
 \includegraphics[width=0.8\textwidth]{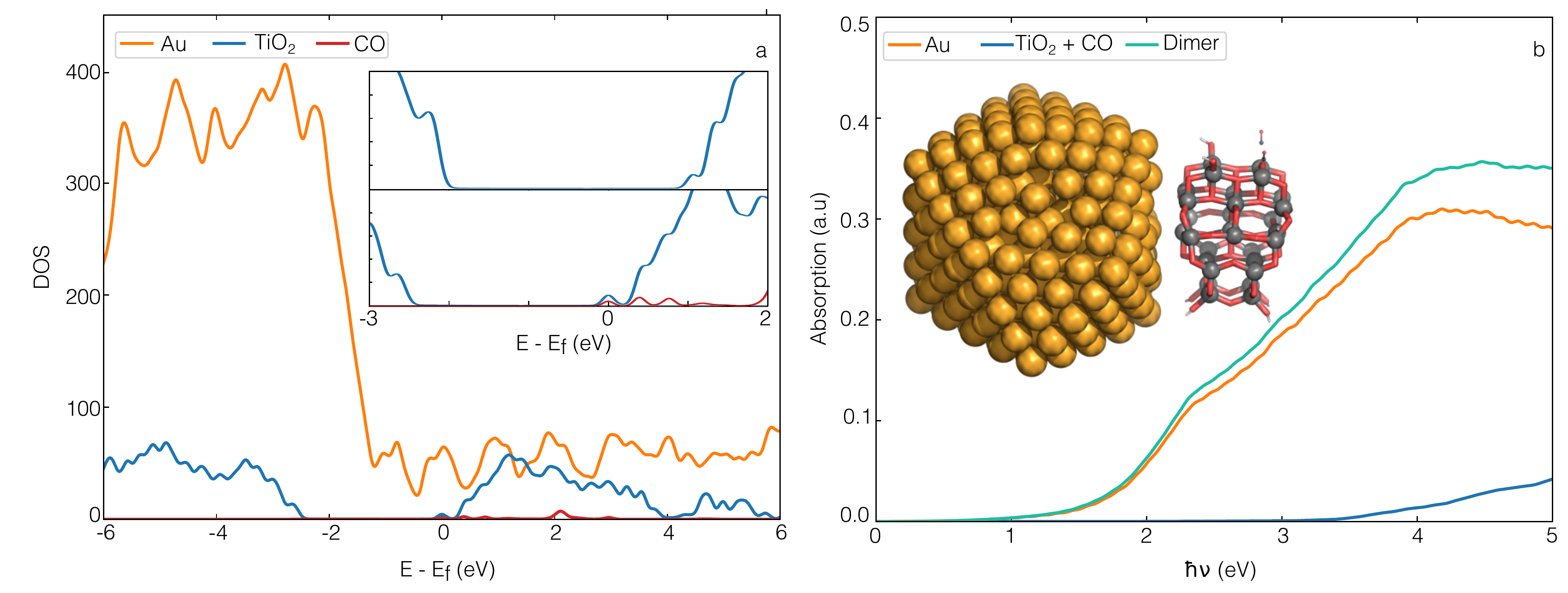}
 \caption{Electronic structure and optical absorption spectrum. (a) Projected density of states (pDOS) for Au, \tio and CO forming the \est cluster. Inset (a) shows a band gap zoom of the pDOS of \tio with (bottom) and without (top) CO molecules. (b) Absorption spectrum of \est, for comparison the spectra of Au and \tio NPs are also shown. Inset of panel (b) presents the optimized structure obtained for \est cluster.  }
 \label{fgr:EandD}
\end{figure}
 
The optical absorption spectrum and the density of states of the \est structure considered here are presented in figure \ref{fgr:EandD}. It is important to highlight that the electronic structures provided by the DFTB Hamiltonian have proven to be suitable to study the bonded anatase and rutile \tio nanocrystals,\cite{gupta2021using} the dynamics of plasmon-induced hot carriers generation in Au NPs\cite{Oscar2019,interplay2019}, and for predicting the photoinjection mechanisms in Dye-Sensitized \tio solar cells.\cite{Negre2012,Negre2014,OvidoB2012} The \est structure employed along this study is presented in the inset of panel b, it is a dimer formed by a 309 Au atoms icosahedral NP and an anatase \tio NP (99 atoms). Oxygen dangling bonds of the \tio NP were passivated with H atoms except for two positions chosen for the interaction with the CO molecules. The whole structure was re-optimized at the DFTB level (see methods section for details). It can be seen that structural relaxation renders slight deviations from the icosahedral and anatase reticle, tending to optimize the interaction between both NP. These distortions do not affect the electronic structure of both particles, nor the LSPR frequency. The projected density of sates (pDOS) of the particle is shown in figure \ref{fgr:EandD}a. It is important to mention that, although we are dealing with a \tio cluster, the electronic structure of the semiconductor is adequate with a band gap of 2.8 eV and without dangling bonds. The inset of figure \ref{fgr:EandD}a compares the pDOS of \tio with and without the interaction with CO. It can be seen that dangling bonds and the gap size reduction arise only from the interaction with the CO molecules.

The absorption spectrum of \est is shown in figure \ref{fgr:EandD}b with a green line and, for comparison, the isolated spectra of Au NP and \tio NP are displayed with orange and blue lines, respectively. The spectrum of the metallic NP is in good agreement with experimental and other theoretical studies as was discussed in \citealt{Oscar2019}. On the other hand, the \tio spectrum resembles the one recently obtained by \citeauthor{ACSgao} at the RT-TD-DFT level of theory for a four-layer slab of \tio \cite{ACSgao} and those reported by \citealt{Fuertes2013} for \tio NPs. The \est spectrum presents a dipolar LSPR that peaks at 2.4 eV and a broad optical absorption band that spreads throughout all the UV-visible region associated with the interband transitions of both NPs. Note that, although the overlap between absorption bands is almost negligible (particularly at the LSPR frequency), a slight broadening and a red-shift (0.1 eV) can be clearly observed for the \est spectrum. This effect arises from the electronic coupling between both NPs and increases as the energy increases due to the coupling of the electronic structure of both systems.               

\begin{figure}
 \centering
 \includegraphics[width=0.8\textwidth]{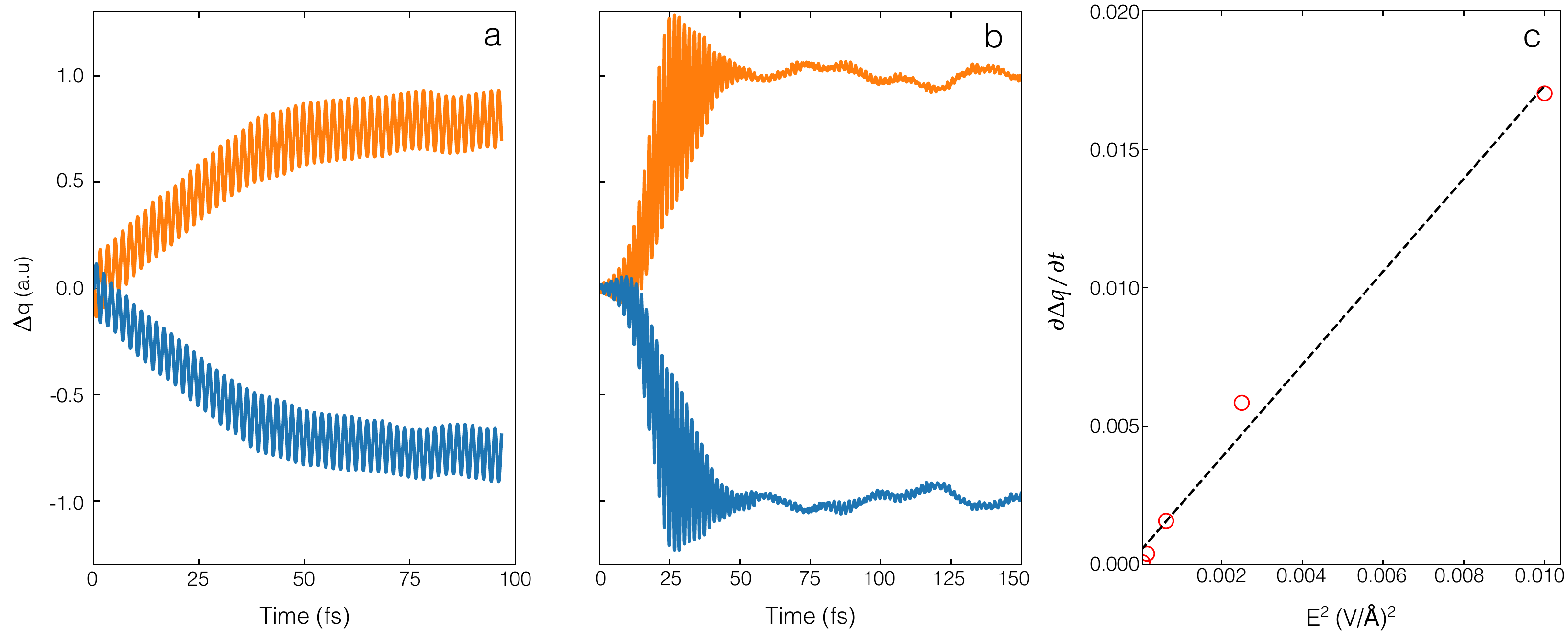}
 \caption{(a) and (b) present the charge evolution ($\Delta$q) of Au NP (orange) and [\tio + CO] (blue), compared to the ground state, as a function of time for a continuum and a pulsed laser, respectively. Positive/negative values of $\Delta$q stand for electron loos/accumulation. Panel (c) presents the charge current extracted from the linear portion of the $\Delta$q graph (panel a) as a function of the laser power}
 \label{fgr:qsvst}
\end{figure}
 
To obtain information on the charge injection dynamics from the metal to the semiconductor, the \est cluster was irradiated with a continuum laser and a 50 fs laser pulse in resonance with the LSPR frequency. The charge separation is described by splitting the structure into two moieties, the metallic NP and the \tio plus the CO molecules. Panels a and b of figure \ref{fgr:qsvst} present the charge evolution of each moiety ($\Delta$q) as a function of time for both kinds of perturbations, note that positive and negative value of $\Delta$q indicates light-induced electron loss or accumulation, respectively. These results show that, under this conditions, almost one electron is transferred from the metal NP to the TiO$_{2}$. From the continuum irradiation condition (panel a) we can estimate the lifetime of the charge injection to be in the order of 50 fs, in close agreement with several experimental studies.\cite{du2013ultrafast,Furube-jacs} Besides, panel c of figure \ref{fgr:qsvst} presents the charge current extracted from the linear portion of the $\Delta$q graph as a function of the laser power. This follows a linear trend, also in agreement with experimental results.\cite{du2013ultrafast,Furube-jacs}  It is important to highlight that the correct description of these two features, the lifetime of the process and the charge transfer dependence with the laser power, validates the Hamiltonian parameterization, the level of theory and the structural model employed.  

The linearity observed in figure \ref{fgr:qsvst}c is an emergent of the dynamics induced by the plasmon energy absorption in Au NPs. The plasmon decay induced by Landau damping in the metal directly
excites electrons to acceptor states in the semiconductor and left
holes in the metal.\cite{Wei2018} It is important to stress that this direct transfer mechanism, mediated by the redistribution of the absorbed energy, is very similar to the type-II charge transfer mechanism in dye-sensitized semiconductors, where is well established that the charge current follows a linear trend with the laser power.\cite{OvidoB2012} Note that the main difference between both mechanisms is that in the dye-sensitized systems laser excites a charge transfer band, whereas, in the hot carrier transfer,  the energy is absorbed by the plasmonic excitation and the direct transfer is an outcome of the energy deactivation.  

\begin{figure}
 \centering
 \includegraphics[width=1\textwidth]{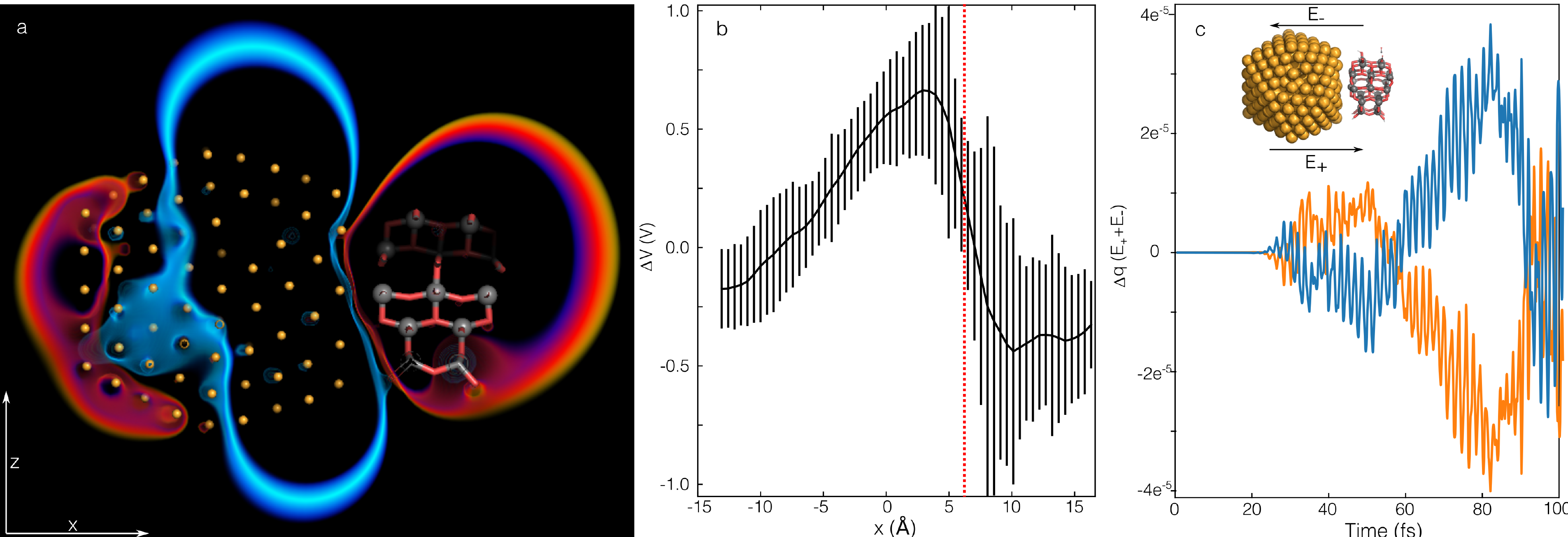}
 \caption{Developed electrostatic potential respect to the ground state $\Delta V = V_{t}-V_{t=0}$; note that this is different from the effective potential an electron would probe. Panel (a) shows a slice of the 3d volume rendering centered at the interparticle axes. Blue tones depict $\Delta V >0$ while violet to yellow scale encodes $\Delta V <0$.  Panel (b) presents $\Delta V$ along $x$ axes, the error bar represents the standard deviation of $\Delta V$ in the $y-z$ plane. The red dotted line helps to visualize the particle interface. (c) Charge rectification obtained when an electric field, polarized in the positive or negative $x$ direction ($E_+$ and $E_-$, respectively), is applied to the system after the charge injection. Orange and blue lines for Au NP and [\tio + CO]  particles, respectively. Note that a negative/positive value indicates that electrons flow preferentially in the negative/positive $x$ direction for the Au NP, and in the positive/negative direction for the \tio NP, respectively }
 \label{fgr:pot}
\end{figure}

To better understand the charge injection process, the supporting figure S1 presents the $\Delta$q obtained with and without the CO under continuum and pulsed irradiation. Several distinctive features can be observed from figures \ref{fgr:qsvst} and S1: (I) Charge separation is stable, namely, charge sloshing between NPs is not observed as it usually happens when big amounts of charge is transferred in simulations without dissipation.  (II) When irradiation ends (the case of a laser pulse), we do not see any trace of back transfer or charge recombination. This result is in agreement with the observation of ps and ns hot-carrier lifetimes in \tio after being transferred from Au.\cite{catal10080916,time-resolved-xray,capturing_long_lived} (III) The presence of the CO increases both the speed of the process and the amount of charge transferred (figure S1). 

Observations (I) and (II) may be easily understood when considering electron-electron, electron-phonon scattering, and energy dissipation to the environment. Nevertheless, none of these are taken into account in results presented in figures \ref{fgr:qsvst} and S1. Hence, the dynamical outcome of our simulations unveils the existence of purely electronic ingredients that stabilize the photoinduced electron-hole pair formed between Au and \tio NPs.  

The first ingredient is the development of a dynamical contribution to the Schottky barrier (DSB) that arises as a consequence of hot carrier injection, locking the back transfer process even in the case of continuum irradiation. The electrostatic contribution to the DSB is presented in figure \ref{fgr:pot}a as a volume rendering of $\Delta V = V_{t}-V_{t=0}$ i.e. the developed electrostatic potential with respect to the ground state; note that this is different from the effective potential an electron would probe. $\Delta V$ is obtained from the dynamical atomic charges, the reference potential being zero at infinity as the system is an isolated cluster. For clarity, we present a slice of the 3d rendering centered at the interparticle axes. In blue tones, the color scale encodes the increase of the electrostatic potential ($\Delta V >0$), and $\Delta V <0$ is represented with a violet to yellow scale. Figure \ref{fgr:pot}a presents $\Delta V$ at the final step of the dynamics while the supporting figure S2 presents several snapshots highlighting the main features of the dynamical  process. Furthermore, figure \ref{fgr:pot}b presents $\Delta V$ along the interparticle axes ($x$ axes), the error bar represents the standard deviation of $\Delta V$ in the $y-z$ plane and the red dotted line helps to localize the particle interface. As it can be clearly seen from figure \ref{fgr:pot} a and b, while the charge injection occurs, the electrostatic component of the point contact potential is distorted as a signature of the charge localization at the interface described previously,\cite{ACSgao} developing a contribution to the DSB localized at the interface of both NPs. The dynamical picture provided by figure S2 shows that during the first femtoseconds the charge (and potential) bounces between both nanoparticles as a consequence of pure plasmonic behavior. As charge leaks to the semiconductor, this oscillatory feature of the potential is lessened by the emergence of a permanent negative potential lobe located over the semiconductor (13fs to 34 fs). After this point, a clear potential interface arises between both NPs while the oscillatory component still can be observable because the continuum laser driving (see $\Delta$q oscillation at the end of the dynamics). This process could be interpreted as the dynamical formation of an electron-hole bound pair where each particle is localized at the Au and \tio NPs, respectively. This purely electronic feature, arising as an outcome of the injection dynamics has been overlooked and can be a missing piece in the complex puzzle that determines the hot carrier lifetime in nanosized structures. 

To add more information regarding the origin of the DSB component, we employed a dimer of the same Au NP and a 852-atom \tio NP, this allows differentiating the bulk and out-layer atoms. The dimer is shined by a laser pulse that produces the charge transfer and we focus on the atomic charge evolution. We compute the difference of the atomic charge at the ground state and at the end of the dynamics ($\Delta q$). Figure S3 presents an $x-z$ projection of $\Delta q$ 50 fs after the laser pulse ends. It can be clearly seen that the injected charge and the remaining holes lock at the interface in good agreement with reference \citenum{ACSgao}. This information points to the origin of DSB as follows. The SB arises because of the Fermi level equalization between both materials. In that process, a ground state charge transfer occurs, and most of that charge pins to the interface giving rise to the SB. Now, if the plasmon-induced charge transfer also pins at the interface, it is expected that the resulting barrier should be modified, giving rise to DSB.

One SB feature is the current rectification provided by the barrier asymmetry. In figure \ref{fgr:pot}c we show the charge rectification that arises when an electric field, polarized in the positive or negative $x$ direction ($E_+$ and $E_-$, respectively), is applied for 25 fs to the system after the charge injection, i.e. 200 fs after the laser pulse. Orange and blue lines present the charge difference obtained for Au NP and [\tio + CO]  particles, respectively. Note that a zero value of  [$\Delta q(E_{+}+E_{-})$] indicates that the barrier does not rectify the electron flow, while a negative/positive value indicates that electrons flow preferentially in the negative/positive $x$ direction for the Au NP, and in the positive/negative direction for the \tio NP, respectively. Hence, the results presented in figure \ref{fgr:pot}c indicate that the DSB rectifies the charge flow as expected for the ground state SB. To compare, figure S4 presents $\Delta$V and the obtained charge rectification at the ground state. It is interesting to note that the rectification obtained after charge injection is one order of magnitude higher than the one obtained at the ground state. This is a clear signature of the rising of a new contribution to the barrier.  


It is important to note that we can not evaluate the strength of this new source of stabilization relative to others like electron-electron scattering, electron-phonon scattering, and energy dissipation to the environment. Nevertheless, the segmentation of the time scales of each process is quite well understood, in fact, that is the key of the famous ``two-temperature model". 1) Hot carriers (non-thermal distribution) are thermalized by the electron-electron scattering within the first tens to hundreds of fs. 2)  The lattice is heated by the electron-phonon scattering in the firsts ps and then, 3) The energy is transferred to the environment in tens of ps. This information has been collected by many experimentalist groups during the last decades and has been reviewed recently by G. Hartland in reference \citenum{2011Hartland}. Because the DBS development follows the charge injection pace, the stabilization provided by it is active orders of magnitude faster than the mechanisms involving phonon scattering, hence we can state confidently that all the mechanisms that involve phonon activation can be safely neglected during the time spanned by our simulations. Even if this time scales proves to be wrong, i.e, if the electron-phonon activates in the first fs of the charge transfer process (we know this is not the case as is discussed in the next paragraphs on the basis of electron-ion dynamics), this does not invalidate the fact that this new contribution to the stabilization of the hot carriers in the \tio is present and has been overlooked due to the difficulty of carrying out dynamical simulations including all the relevant process involved. Furthermore, as we are going to discuss in the following paragraphs, the locking of the back transfer provided by the potential barrier can help to understand the catalytic activity of the injected hot carriers when electron and ions dynamics are coupled.

The second ingredient is an entropic contribution to the charge separation stabilization. As the charge is injected from the Au NP to the \tio the total electronic entropy increases (see figure S5). Therefore,  without additional driving forces, the electron transfer is irreversible once the illumination turns off. A posterior electronic redistribution can occur only if the entropy keeps growing, a process that by construction can not get the system to the initial charge distribution. 

On the other hand, observation (III) points directly to the catalytic effect of the injected hot carriers. It is clear from the density of states (see figure \ref{fgr:EandD}) that the increase of the charge transfer and the speed-up of the process arises as a consequence of the p-doping effect of the CO states near the Fermi level, these extra states promote the charge pumping. In this sense, it is interesting to note that the CO molecules are playing simultaneously the role of the dopant and the reactant. To obtain a realistic picture of the catalytic effect of the hot carrier in the dissociation of the CO molecules it is crucial to account for the dynamics of the atoms. Therefore, we performed simulations of the irradiation where the ion and electron dynamics are coupled as described in the methods section. The supporting figure S6 shows that none of the features described so far regarding the charge injection process are affected by the nuclear motion.

\begin{figure}
 \centering
 \includegraphics[width=0.8\textwidth]{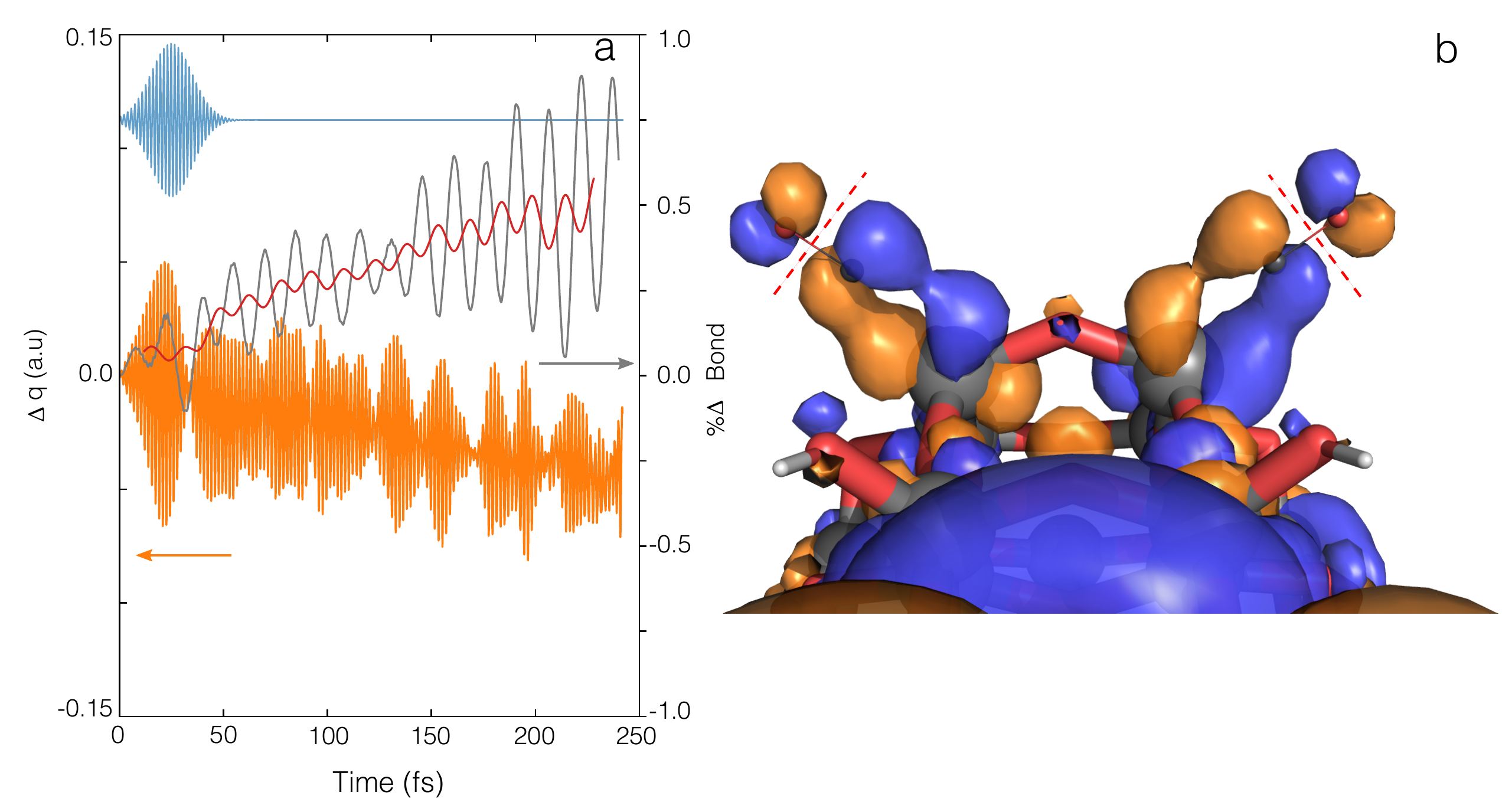}
 \caption{Hot carriers catalytic effect described by coupled electron-ion dynamics.  The \est cluster was irradiated by a laser pulse (blue line panel a). Panel (a) presents the $\Delta q$ for the CO molecules (orange line) and the change of the C-O bond distance as a percentage of the ground state distance (gray line). With a red line, the nearly linear behavior of the bond distance is highlighted with a running average of the raw data. Panel (b) shows a  molecular orbital representative of those being populated during the charge injection. A red dashed line highlights the C-O antibondig character of this orbital. }
 \label{fgr:cat}
\end{figure}

Figure \ref{fgr:cat} presents the results obtained for the CO molecules when considering electron-ion dynamics and a laser pulse irradiation. As a reference, the blue line in panel (a) shows the pulse perturbation, and  $\Delta q$ for both CO molecules is presented with an orange line. It is interesting to note that due to the perturbation, part of the charge received by the \tio is driven to the CO molecules. Remarkably, after the perturbation ends the charge leaking to the CO molecules remains linear until 250 fs, the longest time affordable for this kind of simulation and system size. As the charge reaches the CO molecules, the C-O bond is weakened and the bond distance increases as can be seen with a gray line. The bond distance is informed as a percent of change of the ground state distance. As the bond is weakened the interatomic oscillation increases rapidly keeping also a nearly linear growth pace as the running average (red line) highlights. The elongation of the molecular bond can be rationalized if we look at the molecular orbitals involved in the dynamics. Panel (b) presents a representative molecular orbital of those being populated during the charge injection. As can be appreciated, this possesses an antibonding nature in the C-O coordinate (see nodal plane highlighted with reds dashed lines) which denotes that the injected charge will lead to a photocatalytic dissociation. Figure S7 show that the general trends described for the CO molecules are also obtained for H$_{2}$ molecules, which point to the fact that the conclusions obtained regarding the DSB and its implication in the electron-hole stabilization and photocatalytic effect of the hot carriers are not a particularity of the model system. 

It is important to stress that the results here discussed were obtained with zero initial velocities in all nuclear degrees of freedom (zero phononic temperature). Although this can be an unfavorable condition to observe a photocatalytic process as complex as the one here described, it allows us to highlights the importance of the stabilization of the electron-hole separation furnished by the DSB described earlier. This electronic feature provides the time needed to populate antibonding orbitals of the molecule and to generate the coherent oscillation of the atoms leading to an increase of bond distance,  which is in fact the least likely process, and that will relentlessly lead to the photocatalytic dissociation of the molecule. 

\section{Conclusions}

The exploitation of plasmon-induced hot carriers promises to open new avenues for the development of clean energies and chemical catalysis. The extraction of the carriers before thermalization and the hindrance of carrier recombination is of primordial importance in order to obtain appealing conversion yields. Here we studied the hot carrier injection in the paradigmatic Au-TiO$_2$ system by means of electronic and electron-ion dynamics. In order to simulate nanosized structures, a new DFTB parameterization was presented that successfully describes the electronic properties of the hybrid material. Regarding the charge injection, our simulations show that pure electronic features (without many-body interactions or dissipation to the environment) are enough to justify the electron-hole separation stability. This result unveiled the existence of a dynamical contribution to the Schottky barrier that arises at the charge injection pace, locking the electronic back transfer. This dynamic contribution to the electron-hole stabilization has been overlooked and can be a missing piece in the complex puzzle that determines the hot carrier lifetime in nanosized structures. It is important to highlight that this contribution is active orders of magnitude faster than the electron-phonon scattering and energy dissipation to the environment mechanisms. 

Furthermore, the electron-hole stabilization provided by the DSB is important to describe the photocatalytic effect of the plasmon-induced hot carriers. Here we show that charge leaks to capping molecules placed over the TiO$_{2}$ surface triggering a coherent oscillation of the nuclei that will lead to a photocatalytic dissociation. We expect that our results will add new perspectives to the interpretation of the already detected long-lived hot carrier lifetimes, their catalytical effect, and concomitantly to their technological applications.

\section{Methods}

{\bf{Density Functional Theory}}. 
Density functional calculations were performed using the periodic open source Quantum Espresso software.\cite{QE-2009} The PBE generalized gradient approximation functional\cite{PBE} for the exchange-correlation term was adopted using the high-throughput GBRV pseudopotentials,\cite{GBRV} with standard kinetic energy cutoffs of 40/200~Ry for the wave-function/charge density, respectively. 
10~\AA \ of vacuum space between nanostructures was used in each dimension to avoid spurious interaction between replicated cells. The reciprocal space was described within the Gamma approximation. Unless indicated, structures were fully relaxed using a force convergence of 0.01~eV/\AA{}.

{\bf{Hamiltonian parameterization}}. In order to describe the interaction between the gold and the TiO$_2$ nanoparticles within DFTB, we have combined the already existing tiorg \cite{tiorg-set} and auorg \cite{auorg-set} parameter sets. Both sets are based on the mio parameterization set \cite{mio-set} and extend it with the elements Ti and Au, respectively. The combined set incorporates all existing interactions unchanged, and extends them by the missing Au-Ti interaction. The Hamiltonian and overlap integrals of the Au-Ti dimer were calculated using the same compression radii as defined in the original sets. The Au-Ti repulsive interaction was fitted using a Ti$_2$O$_4$ cluster \cite{Dixon2008} with an interacting gold atom, where the distance between the nearest Ti atom of the TiO$_2$ cluster and the Au atom was varied between 2.34~\AA{} and 2.60~\AA{} (between 90\% and 120\% of the equilibrium bond distance). 

The resulting parameterization was tested by comparing the binding energy and the Au-Ti bond distance obtained by DFT simulations.
For this purpose, and considering the computational cost of a DFT geometry optimization, a model structure was generated using half of a 147 atom icosahedral Au NP and the \tio NP (see figure S8). Furthermore, for the structural optimization, the Au atoms at the cut plane were held fixed. The binding energy difference between the Au and \tio NP computed for the structure optimized at the DFT and the DFTB levels was $\sim$6\% while the Au-Ti bond distance difference for the atoms at the interface was $\sim$1\%    

{\bf{Nanoparticle structure}}. The metallic NP was modeled by an icosahedron of 309 Au atoms. The TiO$_2$ NP was generated using a Wulff construction\cite{WulffConstr} from the anatase bulk geometry.
In this approach we restricted the nanoparticle morphology to that exhibiting the most stable surface (101)\cite{AnataseSurfaces}. As a result, a stoichiometric bipyramid composed of eight (101) facets was built (see figure S9a). The smallest nanoparticle constructed in this manner consists of 105 atoms. The structure presents a danging-bond oxygen atom and low-coordinated Ti atoms at each apex. In order to avoid these sources of structure instability, the former NP was truncated in the apexes, as it has been done in previous studies\cite{bp-model,bp-NP2Bulk}. Finally, six hydrogen atoms and two CO molecules were added to passivate the remaining dangling bonds of oxygen atoms and the low-coordinated Ti atoms, respectively (see figure S9b).
In order to evaluate the structure stability of this nanoparticle, a molecular dynamic simulations based on a reactive force field\cite{reaxFF} was performed using LAMMPS\cite{LAMMPS}. By means of a Nose-Hoover thermostat in the NVT ensamble, the oxide was first heated from 20~K to 300~K in 20~ps, and then kept at that temperature for the remaining 80~ps of the simulation. No reconstructions were observed. Moreover, the average coordination number (number of first neighbors) for Ti and O remain constant during the entire simulation. Finally, no significant changes were observed in the potential energy. It is then safe to conclude that the oxide nanoparticle is stable. Details are given in the supplementary material (see figure 10). 
At last, the whole system was reoptimized at the DFTB level. In the final structure, the CO-TiO{$_2$} binding energy is on the order of 0.4 eV. As expected, this binding energy is smaller than those reported for perfect TiO{$_2$} surfaces.\cite{cgstio2}

{\bf{Electron and electron-ion dynamics}} For the quantum dynamic simulations of this nanosized system the electronic structure was treated at the self-consistent tight-binding charge density-functional (SCC-DFTB) level of theory. The response to laser illumination was obtained by the real time propagation of the density matrix under the perturbation of the light electric field (TD-SCC-DFTB)   by the integration of the Liouville-von Neumann equation of motion. Furthermore, when considering electron-ion real time dynamic simulations, the atomic degree of freedom where coupled to the electronic ones by the Ehrenfest dynamics as described in reference \citenum{Franco_Ehren}. All dynamic simulations where performed within the DFTB+ package, a detailed description of all implemented methods can be found in reference \citenum{DFTBplus}.  To obtain the electronic spectrum from the time dependent dipole moment, the signal was exponentially damped with a time constant of 7 fs. If not otherwise specified the  energy of the pulse or laser irradiation was  2.4 eV. The laser field for the continuum irradiation was set to 0.1 V/\textup{\r{A}} and the pulse laser has a duration of 50 fs and a laser field intensity of 0.3 V/\textup{\r{A}}

\begin{acknowledgement}

The authors thank professor Oscar A. Douglas-Gallardo for inspiring our work and the useful results discussions. M.B and C.G.S acknowledge financial support by Consejo Nacional de Investigaciones Cient\'ificas y T\'ecnicas (CONICET) through Grant PIP 112-2017-0100892CO and PICT-2017-1605. M. B acknowledges financial support by Secretari\'a de Ciencia y Tecnolog\'ia de la Universidad Nacional de C\'ordoba (SECYT-UNC). F.P.B acknowledges the European Union's Horizon 2020 research and innovation programme under the Marie Sklodowska-Curie grant agreement No. 895747. The calculations reported were carried out on computational resources from CCAD Universidad Nacional de C\'ordoba (http://ccad.unc.edu.ar), in particular, the Mendieta and Eulogia clusters. CCAD is part of SNCAD-MinCyT, Rep\'ublica Argentina.

\end{acknowledgement}

\begin{suppinfo}
Structure stability analysis, further information regarding the charge injection, entropy dynamics, and potential barrier dynamics. The catalytic effect over H$_2$ molecules.
\end{suppinfo}

\bibliography{Biblio.bib}

\providecommand{\latin}[1]{#1}
\makeatletter
\providecommand{\doi}
  {\begingroup\let\do\@makeother\dospecials
  \catcode`\{=1 \catcode`\}=2 \doi@aux}
\providecommand{\doi@aux}[1]{\endgroup\texttt{#1}}
\makeatother
\providecommand*\mcitethebibliography{\thebibliography}
\csname @ifundefined\endcsname{endmcitethebibliography}
  {\let\endmcitethebibliography\endthebibliography}{}
\begin{mcitethebibliography}{66}
\providecommand*\natexlab[1]{#1}
\providecommand*\mciteSetBstSublistMode[1]{}
\providecommand*\mciteSetBstMaxWidthForm[2]{}
\providecommand*\mciteBstWouldAddEndPuncttrue
  {\def\EndOfBibitem{\unskip.}}
\providecommand*\mciteBstWouldAddEndPunctfalse
  {\let\EndOfBibitem\relax}
\providecommand*\mciteSetBstMidEndSepPunct[3]{}
\providecommand*\mciteSetBstSublistLabelBeginEnd[3]{}
\providecommand*\EndOfBibitem{}
\mciteSetBstSublistMode{f}
\mciteSetBstMaxWidthForm{subitem}{(\alph{mcitesubitemcount})}
\mciteSetBstSublistLabelBeginEnd
  {\mcitemaxwidthsubitemform\space}
  {\relax}
  {\relax}

\bibitem[Hartland(2011)]{2011Hartland}
Hartland,~G.~V. Optical Studies of Dynamics in Noble Metal Nanostructures.
  \emph{Chem. Rev.} \textbf{2011}, \emph{111}, 3858--3887\relax
\mciteBstWouldAddEndPuncttrue
\mciteSetBstMidEndSepPunct{\mcitedefaultmidpunct}
{\mcitedefaultendpunct}{\mcitedefaultseppunct}\relax
\EndOfBibitem
\bibitem[Hartland \latin{et~al.}(2017)Hartland, Besteiro, Johns, and
  Govorov]{Hartland2017}
Hartland,~G.~V.; Besteiro,~L.~V.; Johns,~P.; Govorov,~A.~O. What’s so Hot
  about Electrons in Metal Nanoparticles? \emph{ACS Energy Lett.}
  \textbf{2017}, \emph{2}, 1641--1653\relax
\mciteBstWouldAddEndPuncttrue
\mciteSetBstMidEndSepPunct{\mcitedefaultmidpunct}
{\mcitedefaultendpunct}{\mcitedefaultseppunct}\relax
\EndOfBibitem
\bibitem[Aslam \latin{et~al.}(2018)Aslam, Rao, Chavez, and Linic]{Linic2018}
Aslam,~U.; Rao,~V.~G.; Chavez,~S.; Linic,~S. Catalytic conversion of solar to
  chemical energy on plasmonic metal nanostructures. \emph{Nat. Catal.}
  \textbf{2018}, \emph{1}, 656\relax
\mciteBstWouldAddEndPuncttrue
\mciteSetBstMidEndSepPunct{\mcitedefaultmidpunct}
{\mcitedefaultendpunct}{\mcitedefaultseppunct}\relax
\EndOfBibitem
\bibitem[Clavero(2014)]{Clavero2014}
Clavero,~C. Plasmon-induced hot-electron generation at nanoparticle/metal-oxide
  interfaces for photovoltaic and photocatalytic devices. \emph{Nat. Photonics}
  \textbf{2014}, \emph{8}, 95--93\relax
\mciteBstWouldAddEndPuncttrue
\mciteSetBstMidEndSepPunct{\mcitedefaultmidpunct}
{\mcitedefaultendpunct}{\mcitedefaultseppunct}\relax
\EndOfBibitem
\bibitem[Brongersma \latin{et~al.}(2015)Brongersma, Halas, and
  Nordlander]{Nordlander2015}
Brongersma,~M.~L.; Halas,~N.~J.; Nordlander,~P. Plasmon-induced hot carrier
  science and technology. \emph{Nat. Nanotechnol.} \textbf{2015}, \emph{10},
  25--34\relax
\mciteBstWouldAddEndPuncttrue
\mciteSetBstMidEndSepPunct{\mcitedefaultmidpunct}
{\mcitedefaultendpunct}{\mcitedefaultseppunct}\relax
\EndOfBibitem
\bibitem[Zhang and Govorov(2014)Zhang, and Govorov]{Govorov2014}
Zhang,~H.; Govorov,~A.~O. Optical Generation of Hot Plasmonic Carriers in Metal
  Nanocrystals: The Effects of Shape and Field Enhancement. \emph{J. Phys.
  Chem. C} \textbf{2014}, \emph{118}, 7606--7614\relax
\mciteBstWouldAddEndPuncttrue
\mciteSetBstMidEndSepPunct{\mcitedefaultmidpunct}
{\mcitedefaultendpunct}{\mcitedefaultseppunct}\relax
\EndOfBibitem
\bibitem[Boerigter \latin{et~al.}(2016)Boerigter, Aslam, and
  Linic]{Boerigter2016}
Boerigter,~C.; Aslam,~U.; Linic,~S. {Mechanism of Charge Transfer from
  Plasmonic Nanostructures to Chemically Attached Materials}. \emph{ACS Nano}
  \textbf{2016}, \emph{10}, 6108--6115\relax
\mciteBstWouldAddEndPuncttrue
\mciteSetBstMidEndSepPunct{\mcitedefaultmidpunct}
{\mcitedefaultendpunct}{\mcitedefaultseppunct}\relax
\EndOfBibitem
\bibitem[Furube \latin{et~al.}(2007)Furube, Du, Hara, Katoh, and
  Tachiya]{Furube-jacs}
Furube,~A.; Du,~L.; Hara,~K.; Katoh,~R.; Tachiya,~M. Ultrafast Plasmon-Induced
  Electron Transfer from Gold Nanodots into TiO2 Nanoparticles. \emph{J. Am.
  Chem. Soc.} \textbf{2007}, \emph{129}, 14852--14853\relax
\mciteBstWouldAddEndPuncttrue
\mciteSetBstMidEndSepPunct{\mcitedefaultmidpunct}
{\mcitedefaultendpunct}{\mcitedefaultseppunct}\relax
\EndOfBibitem
\bibitem[Christopher \latin{et~al.}(2011)Christopher, Xin, and
  Linic]{Christopher2011}
Christopher,~P.; Xin,~H.; Linic,~S. {Visible-light-enhanced catalytic oxidation
  reactions on plasmonic silver nanostructures}. \emph{Nat. Chem.}
  \textbf{2011}, \emph{3}, 467--472\relax
\mciteBstWouldAddEndPuncttrue
\mciteSetBstMidEndSepPunct{\mcitedefaultmidpunct}
{\mcitedefaultendpunct}{\mcitedefaultseppunct}\relax
\EndOfBibitem
\bibitem[Zhao \latin{et~al.}(2018)Zhao, Jin, and Jin]{Zhao_2018}
Zhao,~S.; Jin,~R.; Jin,~R. Opportunities and Challenges in CO2 Reduction by
  Gold- and Silver-Based Electrocatalysts: From Bulk Metals to Nanoparticles
  and Atomically Precise Nanoclusters. \emph{ACS Energy Lett.} \textbf{2018},
  \emph{3}, 452--462\relax
\mciteBstWouldAddEndPuncttrue
\mciteSetBstMidEndSepPunct{\mcitedefaultmidpunct}
{\mcitedefaultendpunct}{\mcitedefaultseppunct}\relax
\EndOfBibitem
\bibitem[Kochuveedu \latin{et~al.}(2012)Kochuveedu, Kim, and
  Kim]{Kochuveedu2012}
Kochuveedu,~S.~T.; Kim,~D.-P.; Kim,~D.~H. Surface-Plasmon-Induced Visible Light
  Photocatalytic Activity of TiO2 Nanospheres Decorated by Au Nanoparticles
  with Controlled Configuration. \emph{J. Phys. Chem. C} \textbf{2012},
  \emph{116}, 2500--2506\relax
\mciteBstWouldAddEndPuncttrue
\mciteSetBstMidEndSepPunct{\mcitedefaultmidpunct}
{\mcitedefaultendpunct}{\mcitedefaultseppunct}\relax
\EndOfBibitem
\bibitem[Mubeen \latin{et~al.}(2013)Mubeen, Lee, Singh, Kr{\"a}mer, Stucky, and
  Moskovits]{mubeen2013}
Mubeen,~S.; Lee,~J.; Singh,~N.; Kr{\"a}mer,~S.; Stucky,~G.~D.; Moskovits,~M. An
  autonomous photosynthetic device in which all charge carriers derive from
  surface plasmons. \emph{Nat. Nanotech.} \textbf{2013}, \emph{8},
  247--251\relax
\mciteBstWouldAddEndPuncttrue
\mciteSetBstMidEndSepPunct{\mcitedefaultmidpunct}
{\mcitedefaultendpunct}{\mcitedefaultseppunct}\relax
\EndOfBibitem
\bibitem[Robatjazi \latin{et~al.}(2015)Robatjazi, Bahauddin, Doiron, and
  Thomann]{robatjazi2015direct}
Robatjazi,~H.; Bahauddin,~S.~M.; Doiron,~C.; Thomann,~I. Direct plasmon-driven
  photoelectrocatalysis. \emph{Nano lett.} \textbf{2015}, \emph{15},
  6155--6161\relax
\mciteBstWouldAddEndPuncttrue
\mciteSetBstMidEndSepPunct{\mcitedefaultmidpunct}
{\mcitedefaultendpunct}{\mcitedefaultseppunct}\relax
\EndOfBibitem
\bibitem[Lee \latin{et~al.}(2012)Lee, Mubeen, Ji, Stucky, and
  Moskovits]{lee2012plasmonic}
Lee,~J.; Mubeen,~S.; Ji,~X.; Stucky,~G.~D.; Moskovits,~M. Plasmonic photoanodes
  for solar water splitting with visible light. \emph{Nano lett.}
  \textbf{2012}, \emph{12}, 5014--5019\relax
\mciteBstWouldAddEndPuncttrue
\mciteSetBstMidEndSepPunct{\mcitedefaultmidpunct}
{\mcitedefaultendpunct}{\mcitedefaultseppunct}\relax
\EndOfBibitem
\bibitem[Mukherjee \latin{et~al.}(2013)Mukherjee, Libisch, Large, Neumann,
  Brown, Cheng, Lassiter, Carter, Nordlander, and Halas]{Halas2013}
Mukherjee,~S.; Libisch,~F.; Large,~N.; Neumann,~O.; Brown,~L.~V.; Cheng,~J.;
  Lassiter,~B.; Carter,~E.~A.; Nordlander,~P.; Halas,~N. Hot Electrons Do the
  Impossible: Plasmon-Induced Dissociation of H2 on Au. \emph{Nano Lett.}
  \textbf{2013}, \emph{13}, 240--247\relax
\mciteBstWouldAddEndPuncttrue
\mciteSetBstMidEndSepPunct{\mcitedefaultmidpunct}
{\mcitedefaultendpunct}{\mcitedefaultseppunct}\relax
\EndOfBibitem
\bibitem[Hattori \latin{et~al.}(2019)Hattori, Abdellah, Meng, Zheng, and
  Sá]{Hattori2019}
Hattori,~Y.; Abdellah,~M.; Meng,~J.; Zheng,~K.; Sá,~J. Simultaneous Hot
  Electron and Hole Injection upon Excitation of Gold Surface Plasmon. \emph{J.
  Phys. Chem. Lett.} \textbf{2019}, \emph{10}, 3140--3146\relax
\mciteBstWouldAddEndPuncttrue
\mciteSetBstMidEndSepPunct{\mcitedefaultmidpunct}
{\mcitedefaultendpunct}{\mcitedefaultseppunct}\relax
\EndOfBibitem
\bibitem[Wu \latin{et~al.}(2015)Wu, Chen, McBride, and Lian]{wu2015efficient}
Wu,~K.; Chen,~J.; McBride,~J.; Lian,~T. Efficient hot-electron transfer by a
  plasmon-induced interfacial charge-transfer transition. \emph{Science}
  \textbf{2015}, \emph{349}, 632--635\relax
\mciteBstWouldAddEndPuncttrue
\mciteSetBstMidEndSepPunct{\mcitedefaultmidpunct}
{\mcitedefaultendpunct}{\mcitedefaultseppunct}\relax
\EndOfBibitem
\bibitem[Du \latin{et~al.}(2013)Du, Furube, Hara, Katoh, and
  Tachiya]{du2013ultrafast}
Du,~L.; Furube,~A.; Hara,~K.; Katoh,~R.; Tachiya,~M. Ultrafast plasmon induced
  electron injection mechanism in gold--TiO 2 nanoparticle system. \emph{J.
  Photochem. Photobiol. C} \textbf{2013}, \emph{15}, 21--30\relax
\mciteBstWouldAddEndPuncttrue
\mciteSetBstMidEndSepPunct{\mcitedefaultmidpunct}
{\mcitedefaultendpunct}{\mcitedefaultseppunct}\relax
\EndOfBibitem
\bibitem[Li \latin{et~al.}(2020)Li, Li, Yang, Ji, Lin, and
  Tomie]{catal10080916}
Li,~B.; Li,~H.; Yang,~C.; Ji,~B.; Lin,~J.; Tomie,~T. Picosecond Lifetime Hot
  Electrons in TiO2 Nanoparticles for High Catalytic Activity. \emph{Catalysts}
  \textbf{2020}, \emph{10}\relax
\mciteBstWouldAddEndPuncttrue
\mciteSetBstMidEndSepPunct{\mcitedefaultmidpunct}
{\mcitedefaultendpunct}{\mcitedefaultseppunct}\relax
\EndOfBibitem
\bibitem[Borgwardt \latin{et~al.}(2020)Borgwardt, Mahl, Roth, Wenthaus,
  Brauße, Blum, Schwarzburg, Liu, Toma, and Gessner]{time-resolved-xray}
Borgwardt,~M.; Mahl,~J.; Roth,~F.; Wenthaus,~L.; Brauße,~F.; Blum,~M.;
  Schwarzburg,~K.; Liu,~G.; Toma,~F.~M.; Gessner,~O. Photoinduced Charge
  Carrier Dynamics and Electron Injection Efficiencies in Au
  Nanoparticle-Sensitized TiO2 Determined with Picosecond Time-Resolved X-ray
  Photoelectron Spectroscopy. \emph{J. Phys. Chem. Lett.} \textbf{2020},
  \emph{11}, 5476--5481\relax
\mciteBstWouldAddEndPuncttrue
\mciteSetBstMidEndSepPunct{\mcitedefaultmidpunct}
{\mcitedefaultendpunct}{\mcitedefaultseppunct}\relax
\EndOfBibitem
\bibitem[Dai \latin{et~al.}(2019)Dai, Jiao, Ma, Liu, Wang, and
  Su]{capturing_long_lived}
Dai,~X.; Jiao,~Z.; Ma,~Z.; Liu,~K.; Wang,~C.; Su,~H. Capturing the Long-Lived
  Photogenerated Electrons in Au/TiO2 upon UV or Visible Irradiation by
  Time-Resolved Infrared Spectroscopy. \emph{J. Phys. Chem. C} \textbf{2019},
  \emph{123}, 20325--20332\relax
\mciteBstWouldAddEndPuncttrue
\mciteSetBstMidEndSepPunct{\mcitedefaultmidpunct}
{\mcitedefaultendpunct}{\mcitedefaultseppunct}\relax
\EndOfBibitem
\bibitem[Zhang \latin{et~al.}(2018)Zhang, He, Guo, Hu, Huang, Mulcahy, and
  Wei]{Wei2018}
Zhang,~Y.; He,~S.; Guo,~W.; Hu,~Y.; Huang,~J.; Mulcahy,~J.~R.; Wei,~W.~D.
  Surface-Plasmon-Driven Hot Electron Photochemistry. \emph{Chem. Rev.}
  \textbf{2018}, \emph{118}, 2927--2954, PMID: 29190069\relax
\mciteBstWouldAddEndPuncttrue
\mciteSetBstMidEndSepPunct{\mcitedefaultmidpunct}
{\mcitedefaultendpunct}{\mcitedefaultseppunct}\relax
\EndOfBibitem
\bibitem[Bernardi \latin{et~al.}(2015)Bernardi, Mustafa, Neaton, and
  Louie]{super-hot1}
Bernardi,~M.; Mustafa,~J.; Neaton,~J.~B.; Louie,~S.~G. Theory and computation
  of hot carriers generated by surface plasmon polaritons in noble metals.
  \emph{Nat. Commun.} \textbf{2015}, \emph{6}, 7044\relax
\mciteBstWouldAddEndPuncttrue
\mciteSetBstMidEndSepPunct{\mcitedefaultmidpunct}
{\mcitedefaultendpunct}{\mcitedefaultseppunct}\relax
\EndOfBibitem
\bibitem[Sundararaman \latin{et~al.}(2014)Sundararaman, Narang, Jermyn,
  Goddard~III, and Atwater]{super-hot2}
Sundararaman,~R.; Narang,~P.; Jermyn,~A.~S.; Goddard~III,~W.~A.; Atwater,~H.~A.
  Theoretical predictions for hot-carrier generation from surface plasmon
  decay. \emph{Nat. Commun.} \textbf{2014}, \emph{5}, 5788\relax
\mciteBstWouldAddEndPuncttrue
\mciteSetBstMidEndSepPunct{\mcitedefaultmidpunct}
{\mcitedefaultendpunct}{\mcitedefaultseppunct}\relax
\EndOfBibitem
\bibitem[Brown \latin{et~al.}(2016)Brown, Sundararaman, Narang, Goddard, and
  Atwater]{super-hot3}
Brown,~A.~M.; Sundararaman,~R.; Narang,~P.; Goddard,~W.~A.; Atwater,~H.~A.
  Nonradiative Plasmon Decay and Hot Carrier Dynamics: Effects of Phonons,
  Surfaces, and Geometry. \emph{ACS Nano} \textbf{2016}, \emph{10},
  957--966\relax
\mciteBstWouldAddEndPuncttrue
\mciteSetBstMidEndSepPunct{\mcitedefaultmidpunct}
{\mcitedefaultendpunct}{\mcitedefaultseppunct}\relax
\EndOfBibitem
\bibitem[Castellanos \latin{et~al.}(2019)Castellanos, Hess, and
  Lischner]{castellanos2019single}
Castellanos,~L.~R.; Hess,~O.; Lischner,~J. Single plasmon hot carrier
  generation in metallic nanoparticles. \emph{Commun. Phys.} \textbf{2019},
  \emph{2}, 47\relax
\mciteBstWouldAddEndPuncttrue
\mciteSetBstMidEndSepPunct{\mcitedefaultmidpunct}
{\mcitedefaultendpunct}{\mcitedefaultseppunct}\relax
\EndOfBibitem
\bibitem[Govorov and Zhang(2015)Govorov, and Zhang]{Govorov2015}
Govorov,~A.~O.; Zhang,~H. Kinetic Density Functional Theory for Plasmonic
  Nanostructures: Breaking of the Plasmon Peak in the Quantum Regime and
  Generation of Hot Electrons. \emph{J. Phys. Chem. C} \textbf{2015},
  \emph{119}, 6181--6194\relax
\mciteBstWouldAddEndPuncttrue
\mciteSetBstMidEndSepPunct{\mcitedefaultmidpunct}
{\mcitedefaultendpunct}{\mcitedefaultseppunct}\relax
\EndOfBibitem
\bibitem[Rossi \latin{et~al.}(2020)Rossi, Erhart, and Kuisma]{Erhart_acsnano}
Rossi,~T.~P.; Erhart,~P.; Kuisma,~M. Hot-Carrier Generation in Plasmonic
  Nanoparticles: The Importance of Atomic Structure. \emph{ACS Nano}
  \textbf{2020}, \emph{14}, 9963--9971\relax
\mciteBstWouldAddEndPuncttrue
\mciteSetBstMidEndSepPunct{\mcitedefaultmidpunct}
{\mcitedefaultendpunct}{\mcitedefaultseppunct}\relax
\EndOfBibitem
\bibitem[Douglas-Gallardo \latin{et~al.}(2016)Douglas-Gallardo, Berdakin, and
  Sánchez]{Oscar2016}
Douglas-Gallardo,~O.~A.; Berdakin,~M.; Sánchez,~C.~G. Atomistic Insights into
  Chemical Interface Damping of Surface Plasmon Excitations in Silver
  Nanoclusters. \emph{J. Phys. Chem. C} \textbf{2016}, \emph{120},
  24389--24399\relax
\mciteBstWouldAddEndPuncttrue
\mciteSetBstMidEndSepPunct{\mcitedefaultmidpunct}
{\mcitedefaultendpunct}{\mcitedefaultseppunct}\relax
\EndOfBibitem
\bibitem[Douglas-Gallardo \latin{et~al.}(2019)Douglas-Gallardo, Berdakin,
  Frauenheim, and Sánchez]{Oscar2019}
Douglas-Gallardo,~O.~A.; Berdakin,~M.; Frauenheim,~T.; Sánchez,~C.~G.
  Plasmon-induced hot-carrier generation differences in gold and silver
  nanoclusters. \emph{Nanoscale} \textbf{2019}, \emph{11}, 8604--8615\relax
\mciteBstWouldAddEndPuncttrue
\mciteSetBstMidEndSepPunct{\mcitedefaultmidpunct}
{\mcitedefaultendpunct}{\mcitedefaultseppunct}\relax
\EndOfBibitem
\bibitem[Berdakin \latin{et~al.}(2020)Berdakin, Douglas-Gallardo, and
  Sánchez]{interplay2019}
Berdakin,~M.; Douglas-Gallardo,~O.~A.; Sánchez,~C.~G. Interplay between Intra-
  and Interband Transitions Associated with the Plasmon-Induced Hot Carrier
  Generation Process in Silver and Gold Nanoclusters. \emph{The Journal of
  Physical Chemistry C} \textbf{2020}, \emph{124}, 1631--1639\relax
\mciteBstWouldAddEndPuncttrue
\mciteSetBstMidEndSepPunct{\mcitedefaultmidpunct}
{\mcitedefaultendpunct}{\mcitedefaultseppunct}\relax
\EndOfBibitem
\bibitem[Manjavacas \latin{et~al.}(2014)Manjavacas, Liu, Kulkarni, and
  Nordlander]{Nordlander2014}
Manjavacas,~A.; Liu,~J.~G.; Kulkarni,~V.; Nordlander,~P. Plasmon-Induced Hot
  Carriers in Metallic Nanoparticles. \emph{ACS Nano} \textbf{2014}, \emph{8},
  7630--7638\relax
\mciteBstWouldAddEndPuncttrue
\mciteSetBstMidEndSepPunct{\mcitedefaultmidpunct}
{\mcitedefaultendpunct}{\mcitedefaultseppunct}\relax
\EndOfBibitem
\bibitem[Román~Castellanos \latin{et~al.}(2020)Román~Castellanos, Kahk, Hess,
  and Lischner]{Lischner_dband}
Román~Castellanos,~L.; Kahk,~J.~M.; Hess,~O.; Lischner,~J. Generation of
  plasmonic hot carriers from d-bands in metallic nanoparticles. \emph{J. Chem.
  Phys.} \textbf{2020}, \emph{152}, 104111\relax
\mciteBstWouldAddEndPuncttrue
\mciteSetBstMidEndSepPunct{\mcitedefaultmidpunct}
{\mcitedefaultendpunct}{\mcitedefaultseppunct}\relax
\EndOfBibitem
\bibitem[Dal~Forno \latin{et~al.}(2018)Dal~Forno, Ranno, and
  Lischner]{Lischner_materials}
Dal~Forno,~S.; Ranno,~L.; Lischner,~J. Material, Size, and Environment
  Dependence of Plasmon-Induced Hot Carriers in Metallic Nanoparticles.
  \emph{J. Phys. Chem. C} \textbf{2018}, \emph{122}, 8517--8527\relax
\mciteBstWouldAddEndPuncttrue
\mciteSetBstMidEndSepPunct{\mcitedefaultmidpunct}
{\mcitedefaultendpunct}{\mcitedefaultseppunct}\relax
\EndOfBibitem
\bibitem[Besteiro \latin{et~al.}(2019)Besteiro, Yu, Wang, Holleitner, Hartland,
  Wiederrecht, and Govorov]{BESTEIRO2019}
Besteiro,~L.~V.; Yu,~P.; Wang,~Z.; Holleitner,~A.~W.; Hartland,~G.~V.;
  Wiederrecht,~G.~P.; Govorov,~A.~O. The fast and the furious: Ultrafast hot
  electrons in plasmonic metastructures. Size and structure matter. \emph{Nano
  Today} \textbf{2019}, \relax
\mciteBstWouldAddEndPunctfalse
\mciteSetBstMidEndSepPunct{\mcitedefaultmidpunct}
{}{\mcitedefaultseppunct}\relax
\EndOfBibitem
\bibitem[Chang \latin{et~al.}(2019)Chang, Besteiro, Sun, Santiago, Gray, Wang,
  and Govorov]{Chang_Le2019}
Chang,~L.; Besteiro,~L.~V.; Sun,~J.; Santiago,~E.~Y.; Gray,~S.~K.; Wang,~Z.;
  Govorov,~A.~O. Electronic Structure of the Plasmons in Metal Nanocrystals:
  Fundamental Limitations for the Energy Efficiency of Hot Electron Generation.
  \emph{ACS Energy Lett.} \textbf{2019}, \emph{4}, 2552--2568\relax
\mciteBstWouldAddEndPuncttrue
\mciteSetBstMidEndSepPunct{\mcitedefaultmidpunct}
{\mcitedefaultendpunct}{\mcitedefaultseppunct}\relax
\EndOfBibitem
\bibitem[Brongersma \latin{et~al.}(2015)Brongersma, Halas, and
  Nordlander]{Brongersma2015}
Brongersma,~M.~L.; Halas,~N.~J.; Nordlander,~P. {Plasmon-induced hot carrier
  science and technology}. \emph{Nat. Nanotech.} \textbf{2015}, \emph{10},
  25--34\relax
\mciteBstWouldAddEndPuncttrue
\mciteSetBstMidEndSepPunct{\mcitedefaultmidpunct}
{\mcitedefaultendpunct}{\mcitedefaultseppunct}\relax
\EndOfBibitem
\bibitem[Zhang \latin{et~al.}(2019)Zhang, Guan, Lischner, Meng, and
  Prezhdo]{nano_Lischner}
Zhang,~J.; Guan,~M.; Lischner,~J.; Meng,~S.; Prezhdo,~O.~V. Coexistence of
  Different Charge-Transfer Mechanisms in the Hot-Carrier Dynamics of Hybrid
  Plasmonic Nanomaterials. \emph{Nano Lett.} \textbf{2019}, \emph{19},
  3187--3193\relax
\mciteBstWouldAddEndPuncttrue
\mciteSetBstMidEndSepPunct{\mcitedefaultmidpunct}
{\mcitedefaultendpunct}{\mcitedefaultseppunct}\relax
\EndOfBibitem
\bibitem[Long and Prezhdo(2014)Long, and Prezhdo]{Prezhdo_instantaneus}
Long,~R.; Prezhdo,~O.~V. Instantaneous Generation of Charge-Separated State on
  TiO2 Surface Sensitized with Plasmonic Nanoparticles. \emph{Journal of the
  American Chemical Society} \textbf{2014}, \emph{136}, 4343--4354\relax
\mciteBstWouldAddEndPuncttrue
\mciteSetBstMidEndSepPunct{\mcitedefaultmidpunct}
{\mcitedefaultendpunct}{\mcitedefaultseppunct}\relax
\EndOfBibitem
\bibitem[Ma and Gao(2019)Ma, and Gao]{ACSgao}
Ma,~J.; Gao,~S. Plasmon-Induced Electron–Hole Separation at the Ag/TiO2(110)
  Interface. \emph{ACS Nano} \textbf{2019}, \emph{13}, 13658--13667\relax
\mciteBstWouldAddEndPuncttrue
\mciteSetBstMidEndSepPunct{\mcitedefaultmidpunct}
{\mcitedefaultendpunct}{\mcitedefaultseppunct}\relax
\EndOfBibitem
\bibitem[Zhang \latin{et~al.}(2018)Zhang, Nelson, Tretiak, Guo, and
  Schatz]{Nelson_acsnano}
Zhang,~Y.; Nelson,~T.; Tretiak,~S.; Guo,~H.; Schatz,~G.~C. Plasmonic
  Hot-Carrier-Mediated Tunable Photochemical Reactions. \emph{ACS Nano}
  \textbf{2018}, \emph{12}, 8415--8422, PMID: 30001116\relax
\mciteBstWouldAddEndPuncttrue
\mciteSetBstMidEndSepPunct{\mcitedefaultmidpunct}
{\mcitedefaultendpunct}{\mcitedefaultseppunct}\relax
\EndOfBibitem
\bibitem[Hull \latin{et~al.}(2020)Hull, Lingerfelt, Li, and Aikens]{Aikens_N2}
Hull,~O.~A.; Lingerfelt,~D.~B.; Li,~X.; Aikens,~C.~M. Electronic Structure and
  Nonadiabatic Dynamics of Atomic Silver Nanowire–N2 Systems. \emph{J. Phys.
  Chem. C} \textbf{2020}, \emph{124}, 20834--20845\relax
\mciteBstWouldAddEndPuncttrue
\mciteSetBstMidEndSepPunct{\mcitedefaultmidpunct}
{\mcitedefaultendpunct}{\mcitedefaultseppunct}\relax
\EndOfBibitem
\bibitem[Yan \latin{et~al.}(2016)Yan, Wang, and Meng]{Sheng_ACS}
Yan,~L.; Wang,~F.; Meng,~S. Quantum Mode Selectivity of Plasmon-Induced Water
  Splitting on Gold Nanoparticles. \emph{ACS Nano} \textbf{2016}, \emph{10},
  5452--5458, PMID: 27127849\relax
\mciteBstWouldAddEndPuncttrue
\mciteSetBstMidEndSepPunct{\mcitedefaultmidpunct}
{\mcitedefaultendpunct}{\mcitedefaultseppunct}\relax
\EndOfBibitem
\bibitem[Medrano and Sánchez(2018)Medrano, and Sánchez]{charly_trap}
Medrano,~C.~R.; Sánchez,~C.~G. Trap-Door-Like Irreversible Photoinduced Charge
  Transfer in a Donor–Acceptor Complex. \emph{J. Phys. Chem. Lett.}
  \textbf{2018}, \emph{9}, 3517--3524\relax
\mciteBstWouldAddEndPuncttrue
\mciteSetBstMidEndSepPunct{\mcitedefaultmidpunct}
{\mcitedefaultendpunct}{\mcitedefaultseppunct}\relax
\EndOfBibitem
\bibitem[Gupta \latin{et~al.}(2021)Gupta, Aradi, Kweon, Keilbart, Goldman,
  Frauenheim, and Kullgren]{gupta2021using}
Gupta,~V.~K.; Aradi,~B.; Kweon,~K.; Keilbart,~N.; Goldman,~N.; Frauenheim,~T.;
  Kullgren,~J. Using DFTB to Model Photocatalytic Anatase-Rutile TiO$_2$
  Nanocrystalline Interfaces and their Band Alignment. 2021\relax
\mciteBstWouldAddEndPuncttrue
\mciteSetBstMidEndSepPunct{\mcitedefaultmidpunct}
{\mcitedefaultendpunct}{\mcitedefaultseppunct}\relax
\EndOfBibitem
\bibitem[Negre \latin{et~al.}(2012)Negre, Fuertes, Oviedo, Oliva, and
  Sánchez]{Negre2012}
Negre,~C. F.~A.; Fuertes,~V.~C.; Oviedo,~M.~B.; Oliva,~F.~Y.; Sánchez,~C.~G.
  Quantum Dynamics of Light-Induced Charge Injection in a Model
  Dye–Nanoparticle Complex. \emph{J. Phys. Chem. C} \textbf{2012},
  \emph{116}, 14748--14753\relax
\mciteBstWouldAddEndPuncttrue
\mciteSetBstMidEndSepPunct{\mcitedefaultmidpunct}
{\mcitedefaultendpunct}{\mcitedefaultseppunct}\relax
\EndOfBibitem
\bibitem[Negre \latin{et~al.}(2014)Negre, Young, Oviedo, Allen, Sánchez,
  Jarzembska, Benedict, Crabtree, Coppens, Brudvig, and Batista]{Negre2014}
Negre,~C. F.~A.; Young,~K.~J.; Oviedo,~M.~B.; Allen,~L.~J.; Sánchez,~C.~G.;
  Jarzembska,~K.~N.; Benedict,~J.~B.; Crabtree,~R.~H.; Coppens,~P.;
  Brudvig,~G.~W.; Batista,~V.~S. Photoelectrochemical Hole Injection Revealed
  in Polyoxotitanate Nanocrystals Functionalized with Organic Adsorbates.
  \emph{J. Am. Chem. Soc.} \textbf{2014}, \emph{136}, 16420--16429\relax
\mciteBstWouldAddEndPuncttrue
\mciteSetBstMidEndSepPunct{\mcitedefaultmidpunct}
{\mcitedefaultendpunct}{\mcitedefaultseppunct}\relax
\EndOfBibitem
\bibitem[Oviedo \latin{et~al.}(2012)Oviedo, Zarate, Negre, Schott,
  Arratia-Pérez, and Sánchez]{OvidoB2012}
Oviedo,~M.~B.; Zarate,~X.; Negre,~C. F.~A.; Schott,~E.; Arratia-Pérez,~R.;
  Sánchez,~C.~G. Quantum Dynamical Simulations as a Tool for Predicting
  Photoinjection Mechanisms in Dye-Sensitized TiO2 Solar Cells. \emph{J. Phys.
  Chem. Lett.} \textbf{2012}, \emph{3}, 2548--2555\relax
\mciteBstWouldAddEndPuncttrue
\mciteSetBstMidEndSepPunct{\mcitedefaultmidpunct}
{\mcitedefaultendpunct}{\mcitedefaultseppunct}\relax
\EndOfBibitem
\bibitem[Fuertes \latin{et~al.}(2013)Fuertes, Negre, Oviedo, Bonafé, Oliva,
  and Sánchez]{Fuertes2013}
Fuertes,~V.~C.; Negre,~C. F.~A.; Oviedo,~M.~B.; Bonafé,~F.~P.; Oliva,~F.~Y.;
  Sánchez,~C.~G. A theoretical study of the optical properties of
  nanostructured TiO2. \emph{Journal of Physics: Condensed Matter}
  \textbf{2013}, \emph{25}, 115304\relax
\mciteBstWouldAddEndPuncttrue
\mciteSetBstMidEndSepPunct{\mcitedefaultmidpunct}
{\mcitedefaultendpunct}{\mcitedefaultseppunct}\relax
\EndOfBibitem
\bibitem[Giannozzi \latin{et~al.}(2009)Giannozzi, Baroni, Bonini, Calandra,
  Car, Cavazzoni, Ceresoli, Chiarotti, Cococcioni, Dabo, {Dal Corso},
  de~Gironcoli, Fabris, Fratesi, Gebauer, Gerstmann, Gougoussis, Kokalj,
  Lazzeri, Martin-Samos, Marzari, Mauri, Mazzarello, Paolini, Pasquarello,
  Paulatto, Sbraccia, Scandolo, Sclauzero, Seitsonen, Smogunov, Umari, and
  Wentzcovitch]{QE-2009}
Giannozzi,~P.; Baroni,~S.; Bonini,~N.; Calandra,~M.; Car,~R.; Cavazzoni,~C.;
  Ceresoli,~D.; Chiarotti,~G.~L.; Cococcioni,~M.; Dabo,~I.; {Dal Corso},~A.;
  de~Gironcoli,~S.; Fabris,~S.; Fratesi,~G.; Gebauer,~R.; Gerstmann,~U.;
  Gougoussis,~C.; Kokalj,~A.; Lazzeri,~M.; Martin-Samos,~L. \latin{et~al.}
  QUANTUM ESPRESSO: a modular and open-source software project for quantum
  simulations of materials. \emph{Journal of Physics: Condensed Matter}
  \textbf{2009}, \emph{21}, 395502 (19pp)\relax
\mciteBstWouldAddEndPuncttrue
\mciteSetBstMidEndSepPunct{\mcitedefaultmidpunct}
{\mcitedefaultendpunct}{\mcitedefaultseppunct}\relax
\EndOfBibitem
\bibitem[Perdew \latin{et~al.}(1996)Perdew, Burke, and Ernzerhof]{PBE}
Perdew,~J.~P.; Burke,~K.; Ernzerhof,~M. Generalized Gradient Approximation Made
  Simple. \emph{Phys. Rev. Lett.} \textbf{1996}, \emph{77}, 3865--3868\relax
\mciteBstWouldAddEndPuncttrue
\mciteSetBstMidEndSepPunct{\mcitedefaultmidpunct}
{\mcitedefaultendpunct}{\mcitedefaultseppunct}\relax
\EndOfBibitem
\bibitem[Garrity \latin{et~al.}(2014)Garrity, Bennett, Rabe, and
  Vanderbilt]{GBRV}
Garrity,~K.~F.; Bennett,~J.~W.; Rabe,~K.~M.; Vanderbilt,~D. Pseudopotentials
  for high-throughput DFT calculations. \emph{Comput. Mater. Sci.}
  \textbf{2014}, \emph{81}, 446 -- 452\relax
\mciteBstWouldAddEndPuncttrue
\mciteSetBstMidEndSepPunct{\mcitedefaultmidpunct}
{\mcitedefaultendpunct}{\mcitedefaultseppunct}\relax
\EndOfBibitem
\bibitem[Dolgonos \latin{et~al.}(2010)Dolgonos, Aradi, Moreira, and
  Frauenheim]{tiorg-set}
Dolgonos,~G.; Aradi,~B.; Moreira,~N.~H.; Frauenheim,~T. An Improved
  Self-Consistent-Charge Density-Functional Tight-Binding (SCC-DFTB) Set of
  Parameters for Simulation of Bulk and Molecular Systems Involving Titanium.
  \emph{J. Chem. Theory Comput.} \textbf{2010}, \emph{6}, 266--278\relax
\mciteBstWouldAddEndPuncttrue
\mciteSetBstMidEndSepPunct{\mcitedefaultmidpunct}
{\mcitedefaultendpunct}{\mcitedefaultseppunct}\relax
\EndOfBibitem
\bibitem[Fihey \latin{et~al.}(2015)Fihey, Hettich, Touzeau, Maurel, Perrier,
  K\"ohler, Aradi, and Frauenheim]{auorg-set}
Fihey,~A.; Hettich,~C.; Touzeau,~J.; Maurel,~F.; Perrier,~A.; K\"ohler,~C.;
  Aradi,~B.; Frauenheim,~T. SCC-DFTB parameters for simulating hybrid
  gold-thiolates compounds. \emph{J. Comp. Chem.} \textbf{2015}, \emph{36},
  2075\relax
\mciteBstWouldAddEndPuncttrue
\mciteSetBstMidEndSepPunct{\mcitedefaultmidpunct}
{\mcitedefaultendpunct}{\mcitedefaultseppunct}\relax
\EndOfBibitem
\bibitem[Elstner \latin{et~al.}(1998)Elstner, Porezag, Jungnickel, Elsner,
  Haugk, Frauenheim, Suhai, and Seifert]{mio-set}
Elstner,~M.; Porezag,~D.; Jungnickel,~G.; Elsner,~J.; Haugk,~M.;
  Frauenheim,~T.; Suhai,~S.; Seifert,~G. Self-consistent-charge
  density-functional tight-binding method for simulations of complex materials
  properties. \emph{Phys. Rev. B} \textbf{1998}, \emph{58}, 7260--7268\relax
\mciteBstWouldAddEndPuncttrue
\mciteSetBstMidEndSepPunct{\mcitedefaultmidpunct}
{\mcitedefaultendpunct}{\mcitedefaultseppunct}\relax
\EndOfBibitem
\bibitem[Li and Dixon(2008)Li, and Dixon]{Dixon2008}
Li,~S.; Dixon,~D.~A. Molecular Structures and Energetics of the (TiO2)n (n =
  1−4) Clusters and Their Anions. \emph{J. Phys. Chem. A} \textbf{2008},
  \emph{112}, 6646--6666\relax
\mciteBstWouldAddEndPuncttrue
\mciteSetBstMidEndSepPunct{\mcitedefaultmidpunct}
{\mcitedefaultendpunct}{\mcitedefaultseppunct}\relax
\EndOfBibitem
\bibitem[Wulff(1901)]{WulffConstr}
Wulff,~G. On the question of speed of growth and dissolution of crystal
  surfaces. \emph{Z. Kristallogr} \textbf{1901}, \emph{34}, 449\relax
\mciteBstWouldAddEndPuncttrue
\mciteSetBstMidEndSepPunct{\mcitedefaultmidpunct}
{\mcitedefaultendpunct}{\mcitedefaultseppunct}\relax
\EndOfBibitem
\bibitem[Lazzeri \latin{et~al.}(2001)Lazzeri, Vittadini, and
  Selloni]{AnataseSurfaces}
Lazzeri,~M.; Vittadini,~A.; Selloni,~A. Structure and energetics of
  stoichiometric ${\mathrm{TiO}}_{2}$ anatase surfaces. \emph{Phys. Rev. B}
  \textbf{2001}, \emph{63}, 155409\relax
\mciteBstWouldAddEndPuncttrue
\mciteSetBstMidEndSepPunct{\mcitedefaultmidpunct}
{\mcitedefaultendpunct}{\mcitedefaultseppunct}\relax
\EndOfBibitem
\bibitem[Barnard \latin{et~al.}(2006)Barnard, Erdin, Lin, Zapol, and
  Halley]{bp-model}
Barnard,~A.~S.; Erdin,~S.; Lin,~Y.; Zapol,~P.; Halley,~J.~W. Modeling the
  structure and electronic properties of $\mathrm{Ti}{\mathrm{O}}_{2}$
  nanoparticles. \emph{Phys. Rev. B} \textbf{2006}, \emph{73}, 205405\relax
\mciteBstWouldAddEndPuncttrue
\mciteSetBstMidEndSepPunct{\mcitedefaultmidpunct}
{\mcitedefaultendpunct}{\mcitedefaultseppunct}\relax
\EndOfBibitem
\bibitem[Lamiel-Garcia \latin{et~al.}(2017)Lamiel-Garcia, Ko, Lee, Bromley, and
  Illas]{bp-NP2Bulk}
Lamiel-Garcia,~O.; Ko,~K.~C.; Lee,~J.~Y.; Bromley,~S.~T.; Illas,~F. When
  Anatase Nanoparticles Become Bulklike: Properties of Realistic TiO2
  Nanoparticles in the 1–6 nm Size Range from All Electron Relativistic
  Density Functional Theory Based Calculations. \emph{J. Chem. Theory Comput.}
  \textbf{2017}, \emph{13}, 1785--1793\relax
\mciteBstWouldAddEndPuncttrue
\mciteSetBstMidEndSepPunct{\mcitedefaultmidpunct}
{\mcitedefaultendpunct}{\mcitedefaultseppunct}\relax
\EndOfBibitem
\bibitem[Huygh \latin{et~al.}(2014)Huygh, Bogaerts, {van Duin}, and
  Neyts]{reaxFF}
Huygh,~S.; Bogaerts,~A.; {van Duin},~A.~C.; Neyts,~E.~C. Development of a
  ReaxFF reactive force field for intrinsic point defects in titanium dioxide.
  \emph{Comput. Mater. Sci.} \textbf{2014}, \emph{95}, 579--591\relax
\mciteBstWouldAddEndPuncttrue
\mciteSetBstMidEndSepPunct{\mcitedefaultmidpunct}
{\mcitedefaultendpunct}{\mcitedefaultseppunct}\relax
\EndOfBibitem
\bibitem[Plimpton(1995)]{LAMMPS}
Plimpton,~S. Fast Parallel Algorithms for Short-Range Molecular Dynamics.
  \emph{J. Comput. Phys.} \textbf{1995}, \emph{117}, 1–19\relax
\mciteBstWouldAddEndPuncttrue
\mciteSetBstMidEndSepPunct{\mcitedefaultmidpunct}
{\mcitedefaultendpunct}{\mcitedefaultseppunct}\relax
\EndOfBibitem
\bibitem[Liu \latin{et~al.}(2003)Liu, Gong, Kohanoff, Sanchez, and Hu]{cgstio2}
Liu,~Z.-P.; Gong,~X.-Q.; Kohanoff,~J.; Sanchez,~C.; Hu,~P. Catalytic Role of
  Metal Oxides in Gold-Based Catalysts: A First Principles Study of CO
  Oxidation on ${\mathrm{T}\mathrm{i}\mathrm{O}}_{2}$ Supported Au. \emph{Phys.
  Rev. Lett.} \textbf{2003}, \emph{91}, 266102\relax
\mciteBstWouldAddEndPuncttrue
\mciteSetBstMidEndSepPunct{\mcitedefaultmidpunct}
{\mcitedefaultendpunct}{\mcitedefaultseppunct}\relax
\EndOfBibitem
\bibitem[Bonafé \latin{et~al.}(2020)Bonafé, Aradi, Hourahine, Medrano,
  Hernández, Frauenheim, and Sánchez]{Franco_Ehren}
Bonafé,~F.~P.; Aradi,~B.; Hourahine,~B.; Medrano,~C.~R.; Hernández,~F.~J.;
  Frauenheim,~T.; Sánchez,~C.~G. A Real-Time Time-Dependent Density Functional
  Tight-Binding Implementation for Semiclassical Excited State
  Electron–Nuclear Dynamics and Pump–Probe Spectroscopy Simulations.
  \emph{J. Chem. Theory Comput.} \textbf{2020}, \emph{16}, 4454--4469, PMID:
  32511909\relax
\mciteBstWouldAddEndPuncttrue
\mciteSetBstMidEndSepPunct{\mcitedefaultmidpunct}
{\mcitedefaultendpunct}{\mcitedefaultseppunct}\relax
\EndOfBibitem
\bibitem[Hourahine \latin{et~al.}(2020)Hourahine, Aradi, Blum, Bonafé,
  Buccheri, Camacho, Cevallos, Deshaye, Dumitrică, Dominguez, Ehlert, Elstner,
  van~der Heide, Hermann, Irle, Kranz, Köhler, Kowalczyk, Kubař, Lee,
  Lutsker, Maurer, Min, Mitchell, Negre, Niehaus, Niklasson, Page, Pecchia,
  Penazzi, Persson, Řezáč, Sánchez, Sternberg, Stöhr, Stuckenberg,
  Tkatchenko, Yu, and Frauenheim]{DFTBplus}
Hourahine,~B.; Aradi,~B.; Blum,~V.; Bonafé,~F.; Buccheri,~A.; Camacho,~C.;
  Cevallos,~C.; Deshaye,~M.~Y.; Dumitrică,~T.; Dominguez,~A.; Ehlert,~S.;
  Elstner,~M.; van~der Heide,~T.; Hermann,~J.; Irle,~S.; Kranz,~J.~J.;
  Köhler,~C.; Kowalczyk,~T.; Kubař,~T.; Lee,~I.~S. \latin{et~al.}  DFTB+, a
  software package for efficient approximate density functional theory based
  atomistic simulations. \emph{J. Chem. Phys.} \textbf{2020}, \emph{152},
  124101\relax
\mciteBstWouldAddEndPuncttrue
\mciteSetBstMidEndSepPunct{\mcitedefaultmidpunct}
{\mcitedefaultendpunct}{\mcitedefaultseppunct}\relax
\EndOfBibitem
\end{mcitethebibliography}

\end{document}